\documentclass[runningheads]{llncs}
\usepackage{amsmath,graphicx}
\usepackage{amssymb}
\usepackage{xcolor}
\usepackage{diagbox}
\usepackage{booktabs} 
\usepackage{url}
\usepackage[T1]{fontenc}
\usepackage{todonotes}
\usepackage[colorlinks=true,citecolor=teal]{hyperref}
\usepackage{graphicx}

\newcommand{\bs}[1]{\boldsymbol{#1}}

\begin{document}
\title{Patcher: Patch Transformers with Mixture of Experts for Precise Medical Image Segmentation}
\titlerunning{Patcher: Patch Transformers with Mixture of Experts}
\author{Yanglan Ou\inst{1}, Ye Yuan\inst{2}, Xiaolei Huang\inst{1}, Stephen T.C. Wong\inst{3}, \break John Volpi\inst{4}, James Z. Wang\inst{1}, Kelvin Wong\inst{3}}
\institute{The Pennsylvania State University, University Park, Pennsylvania, USA \and
Carnegie Mellon University, Pittsburgh, Pennsylvania, USA \and
TT and WF Chao Center for BRAIN \& Houston Methodist Cancer Center, Houston Methodist Hospital, Houston, Texas, USA \and
Eddy Scurlock Comprehensive Stroke Center, Department of Neurology,\break Houston Methodist Hospital, Houston, Texas, USA}
\authorrunning{Y. Ou et al.}
\maketitle              %
\begin{abstract}
We present a new encoder-decoder Vision Transformer architecture, Patcher, for medical image segmentation. Unlike standard Vision Transformers, it employs Patcher blocks that segment an image into large patches, each of which is further divided into small patches. Transformers are applied to the small patches within a large patch, which constrains the receptive field of each pixel. We intentionally make the large patches overlap to enhance intra-patch communication. The encoder employs a cascade of Patcher blocks with increasing receptive fields to extract features from local to global levels. This design allows Patcher to benefit from both the coarse-to-fine feature extraction common in CNNs and the superior spatial relationship modeling of Transformers. We also propose a new mixture-of-experts (MoE) based decoder, which treats the feature maps from the encoder as experts and selects a suitable set of expert features to predict the label for each pixel. The use of MoE enables better specializations of the expert features and reduces interference between them during inference. Extensive experiments demonstrate that Patcher outperforms state-of-the-art Transformer- and CNN-based approaches significantly on stroke lesion segmentation and polyp segmentation. Code for Patcher is released to facilitate related research.\footnote{Code: \url{https://github.com/YanglanOu/patcher.git}}

\end{abstract}

\section{Introduction}

Deep learning-based medical image segmentation has many important applications in computer-aided diagnosis and treatment. Until recently, the field of medical image segmentation has mainly been dominated by convolutional neural networks (CNNs). U-Net and its variants~\cite{ronneberger2015u,cciccek20163d,zhou2018unet++,oktay2018attention,jha2019resunet++,chen2021transunet} are a representative class of CNN-based models, which are often the preferred networks for image segmentation. These models mainly adopt an encoder-decoder architecture, where an encoder uses a cascade of CNN layers with increasing receptive fields to capture both local and global features while a decoder leverages skip-connections and deconvolutional layers to effectively combine the local and global features into the final prediction. Despite the tremendous success of CNN-based models, their drawbacks start to become apparent as their performance saturates. First, CNNs are suboptimal at modeling global context. While increasing the depth of CNN-based models can enlarge the receptive fields, it also leads to problems such as diminishing feature reuse~\cite{srivastava2015highway}, where low-level features are diluted by consecutive multiplications. Second, the translation invariance of CNNs is a double-edged sword -- it allows CNNs to generalize better, but also severely constrains their ability to reason about the spatial relationships between pixels.

Originally designed for NLP tasks, Transformers~\cite{vaswani2017attention} have recently become popular in vision domains since the invention of ViT~\cite{dosovitskiy2020image}. By design, Vision Transformers have the ability to address the two aforementioned drawbacks of CNNs: (1) They can effectively model the global context by segmenting an image into patches and applying self-attention to them; (2) The use of positional encodings endows Transformers with strong abilities to model spatial relationships between patches. Due to these advantages, many Vision Transformer models have been proposed for image segmentation. For instance, SETR~\cite{zheng2021rethinking} employs a ViT-based encoder in an encoder-decoder architecture to attain superior performance over CNNs. Recently, Swin Transformer~\cite{liu2021Swin} adopts a hierarchical design to improve the efficiency of Vision Transformers and achieves SOTA results on various vision tasks including image segmentation. The success of Vision Transformers has attracted significant attention in the domain of medical image segmentation. For example, Swin-Unet~\cite{cao2021swin}, LambdaUNet~\cite{ou2021lambdaunet}, and U-NetR~\cite{hatamizadeh2022unetr} replace convolutional layers with Transformers in a U-Net-like architecture. Other models like \mbox{TransUNet}~\cite{chen2021transunet}, U-Net Transformer~\cite{petit2021u}, and TransFuse~\cite{zhang2021transfuse} adopt a hybrid approach where they use Transformers to capture global context and convolutional layers to extract local features. So far, most prior work utilizes Transformers mainly to extract patch-level features instead of fine pixel-level features. Given Transformers' strong abilities to model spatial relationships, we believe there is an opportunity to fully leverage Transformers for extracting fine-grained pixel-level features without delegating it to convolutional layers.

To this end, we propose a new encoder-decoder Vision Transformer architecture, Patcher, that uses Transformers for extracting fine-grained local features in addition to global features. Its key component is the Patcher block, which segments an image into large patches (\emph{e.g.,} $32 \times 32$), each of which is further divided into small patches (\emph{e.g.,} $2 \times 2$). Transformers are applied to the small patches within each large patch to extract pixel-level features. Each large patch constrains the receptive fields of the pixels inside, and we intentionally make the large patches overlap to enhance intra-patch communication. The encoder employs a cascade of Patcher blocks with increasing receptive fields to output a sequence of features maps extracted from local to global levels. This design allows Patcher to combine the best of both worlds: it shares the classic coarse-to-fine feature extraction common in CNNs and also enjoys Transformers' superior spatial relationship modeling power to capture low-level details. Moreover, we make the observation that image segmentation models mainly require local features for some pixels (\emph{e.g.,} edge pixels) while relying more on global features for other pixels (\emph{e.g.,} pixels inside a global shape). This motivates us to further propose a new mixture-of-experts (MoE) based decoder. It treats the feature maps from the encoder as experts and learns a gating network to select a suitable set of expert features to predict the label for each pixel. By using MoE, the model can learn more specialized and disentangled expert feature maps and reduce interference between them during inference.

The contributions of this paper are as follows: (1) We propose a new Vision Transformer architecture that can effectively extract fine-grained pixel-level features without relying on convolutional layers;
(2) We propose a new MoE-based decoder that enables better specialization and disentanglement of features maps, which substantially improves performance;
(3) Extensive experiments demonstrate that our method outperforms SOTA Transformer- and CNN-based approaches significantly on stroke lesion segmentation and polyp segmentation.

\vspace{-2mm}
\section{Method}

\begin{figure*}[t]
\centering
\includegraphics[width=\linewidth]{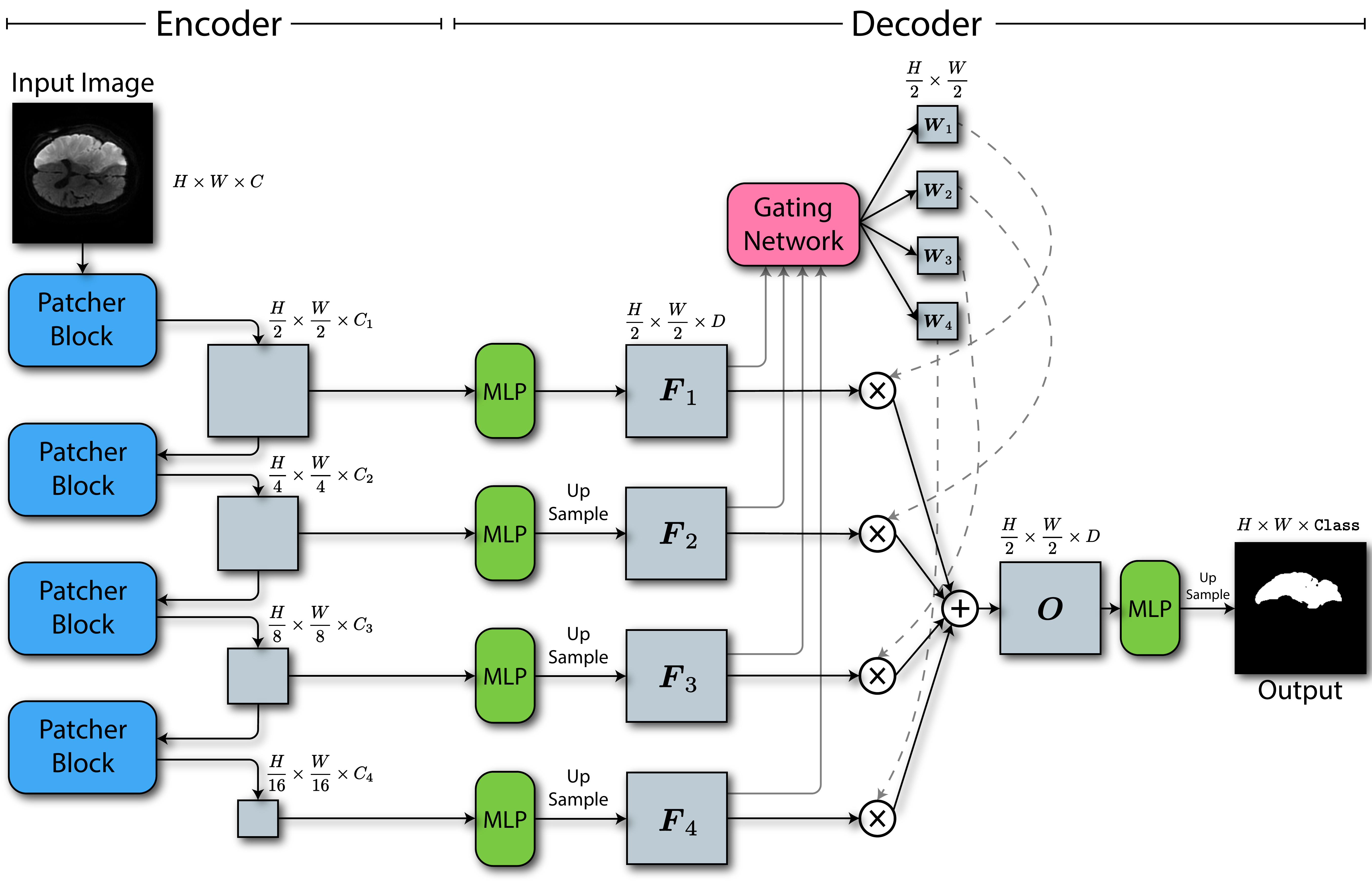}
\vspace{-8mm}
\caption{Model overview of Patcher. The encoder uses a cascade of Patcher blocks to extract expert features from local to global levels. The MoE-based decoder uses a gating network to select a suitable set of expert features for the prediction of each pixel.}
\label{fig:overview}
\vspace{-4mm}
\end{figure*}

\subsection{Overall Architecture}
The network architecture of Patcher is outlined in Fig.~\ref{fig:overview}.
Given an input image of size $H \times W \times C$, Patcher first uses an encoder to extract features from the input image. The encoder contains a cascade of Transformer-based Patcher blocks (detailed in Sec.~\ref{sec:patcher_block}), which produce a sequence of features maps capturing visual features from local to global levels with increasing receptive fields. These feature maps are then input to a decoder with a mixture-of-experts (MoE) design, where each of the feature maps from the encoder serves as an expert. A gating network in the decoder outputs weight maps for the expert feature maps and uses the weights to obtain a combined feature map. A multi-layer perceptron (MLP) and an up-sampling layer are then used to process the combined feature map into the final segmentation output. The MoE-based design increases the specialization of different levels of features while reducing the interference between them. It allows the network to make predictions for each pixel by choosing a suitable set of expert features. For example, the network may need global features for pixels inside a particular global shape, while it may require local features to capture fine details at segmentation boundaries. Finally, we use the standard binary cross-entropy (BCE) loss for image segmentation to train Patcher.

\begin{figure*}[t]
\centering
\includegraphics[width=\linewidth]{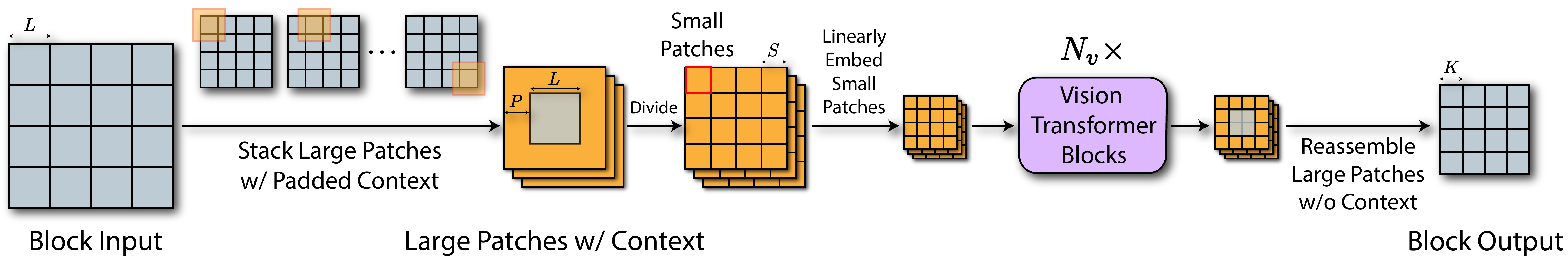}
\vspace{-6mm}
\caption{Overview of the Patcher block. The input is segmented into large patches with overlapping context, each of which is further divided into small patches. The small patches are processed by a sequence of Vision Transformer blocks to extract fine-grained features. The final output is produced by reassembling the large patches.}
\label{fig:patcher_block}
\vspace{-4mm}
\end{figure*}

\subsection{Patcher Block}
\label{sec:patcher_block}

The key component inside the Patcher encoder is the Patcher block, which is a versatile Transformer-based module that can extract visual features from the block input at different spatial levels. The overview of the Patcher block is outlined in Fig.~\ref{fig:patcher_block}. We first divide the block input of spatial dimensions $H \times W$ into an $N_h \times N_w$ grid of large patches. Each of the large patches is of size $L \times L$, where $N_h = H/L$ and $N_w = W/L$. We further pad each large patch with $P$ pixels from neighboring patches on each side, forming a patch of size $(L + 2P) \times (L + 2P)$. We stack the large patches with padded context along the batch dimension (batch size $B = B_0N_hN_w$ where $B_0$ is the original batch size), so different large patches will not interfere with each other in the subsequent operations. Each large patch defines a receptive field similar to the kernel in CNNs, with the difference that all pixels inside the large patch share the same receptive field. Therefore, it is crucial to have the padded context since it enlarges the receptive field of pixels, which is especially important for pixels close to the patch boundaries. Next, we partition the stacked large patches further into an $M_h \times M_w$ grid of small patches, each of size $S \times S$. Similar to prior work, we linearly embed all pixels inside each small patch into a token, and the tokens of all small patches form a sequence. We then use $N_v$ Vision Transformer blocks~\cite{dosovitskiy2020image} to process the sequence to model the relationship between patches and extract useful visual features. Inspired by SegFormer~\cite{xie2021segformer}, we do not use positional encodings but mix convolutional layers inside the MLPs of the Transformer to capture spatial relationships. We also use the efficient self-attention in SegFormer to further reduce the computational cost. We provide more details of the Vision Transformer blocks in Appendix~\ref{appx: ViT}. The output feature maps of the Transformer blocks have a spatial dimension of $M_h \times M_w$ with batch size $B$. We take the center $K \times K$ area from the feature maps where $K = L/S$, which excludes the padded context and corresponds to the actual large patches. We then reassemble the large patches based on their locations in the original image to form the final output with spatial dimensions $\frac{H}{S} \times \frac{W}{S}$. There are two important hyperparameters of the Patcher block: (1) large patch size $L$, which defines the receptive field and allows feature extraction at local or global levels; (2) padded context size $P$, which controls how much information from neighboring large patches are used.

\subsection{Patcher Encoder}
As shown in Fig.~\ref{fig:overview}, the Patcher encoder employs a cascade of four Patcher blocks to produce four feature maps with decreasing spatial dimensions and increasing receptive fields. The small patch size $S$ is set to 2 for all the blocks, which means the spatial dimensions are halved after each block. The large patch size $L$ and padded context size $P$ are set to 32 and 8, respectively. By setting $L$ and $P$ the same for all the blocks, we allow deeper Patcher blocks to have a larger receptive field, therefore gradually shifting the focus of the blocks from capturing local features to global features. This encoder design mirrors the behavior of its CNN counterparts such as \mbox{U-Net}~\cite{ronneberger2015u}, which has been proven effective. Therefore, the Patcher encoder combines the best of both worlds -- it not only benefits from the superior spatial relationship modeling of Transformers but also enjoys the effective coarse-to-fine feature extraction of CNNs.

\subsection{Mixture of Experts Decoder}
The decoder follows an MoE design, where it treats the four features maps from the encoder as experts. As illustrated in Fig.~\ref{fig:overview}, the decoder first uses pixel-wise MLPs to process each of the feature maps and then upsamples them to the size of the first feature maps, \emph{i.e.,} $\frac{H}{2}\times\frac{W}{2}\times D$, where $D$ is the number of channels after the MLPs. We use $[\bs{F}_1, \bs{F}_2, \bs{F}_3, \bs{F}_4]$ to denote the upsampled features, which are also called expert features. Next, a gating network takes the expert features as input and produces the weight maps $[\bs{W}_1, \bs{W}_2, \bs{W}_3, \bs{W}_4]$ for the expert feature maps, where each weight map is of size $\frac{H}{2}\times\frac{W}{2}$. The weight maps sum to 1 per pixel, \emph{i.e.,} $\bs{W}_1 + \bs{W}_2 + \bs{W}_3 + \bs{W}_4 = \bs{1}$. The gating network first concatenates all the expert feature maps along channels and uses several convolutional layers and a final softmax layer to process the concatenated features into the weight maps. We then use the weight maps to produce the combined feature map $\bs{O}$:
\begin{equation}
    \bs{O} = \sum_{i=1}^4\bs{W}_i\odot \bs{F}_i\,,
\end{equation}
where $\odot$ denotes pixel-wise multiplication. The combined feature map $\bs{O}$ then passes through another MLP to predict the segmentation logits before being upsampled to the original image size. The MoE design of the decoder allows the network to learn more specialized feature maps and reduce the interference between them. For the prediction of each pixel, the gating function chooses a suitable set of features by weighing the importance of global \emph{vs.} local features.

\section{Experiments}

\noindent\textbf{Dataset.} We perform experiments on two important medical image segmentation tasks. (1)~\textbf{Stroke lesion segmentation}: We collect 99 acute ischemic stroke cases for this research with help from a major academic hospital. The dataset contains 99 cases (2,451 images) in total. Each image has two channels, the eADC and DWI, from ischemic stroke patients. We use 67 cases (1,652 images) for training, 20 cases (499 images) for validation, and 12 cases (300 images) for testing. The test data has a high diversity of lesion sizes, locations, and stroke types. (2)~\textbf{Polyp  segmentation}: We further conduct experiments on a public dataset of gastrointestinal polyp images, \href{https://datasets.simula.no/kvasir-seg/}{Kvasir-SEG}~\cite{jha2020kvasir}. The dataset contains 1,000 RGB images and corresponding segmentation masks. We randomly split the dataset with a ratio of 8:1:1 for training, validation, and testing.

\vspace{2mm}
\noindent\textbf{Implementation Details.} 
We train our models on two NVIDIA RTX 6000 GPUs using a batch size of 8, which takes around 12 hours. For both stroke lesion and polyp segmentation, the input image is scaled to $256 \times 256$. During training, we randomly scale the image with a ratio from 0.7 to 2.0 and crop the image to $256 \times 256$ again. For the encoder's four Patcher blocks, we use $[64, 128, 320, 512]$ for the Transformer embedding dimensions and $[3,6,40,3]$ for the number of Transformer blocks $N_v$. For the decoder, all the MLPs have hidden dimensions $[256,256]$ with ReLU activations. The expert feature maps $[\bs{F}_1,\bs{F}_2,\bs{F}_3,\bs{F}_4]$ have $D=256$ channels. The gating network uses four $3\times 3$ convolutional layers with channels $[256, 256, 256, 4]$ and ReLU activations. For stroke lesion segmentation, we use the AdamW~\cite{loshchilov2017decoupled} optimizer with an initial learning rate of $6e-5$. For polyp segmentation, we use the Adam~\cite{kingma2014adam} optimizer with an initial learning rate of $1e-4$. We adopt a polynomial-decay learning rate schedule for both datasets. We also use the Intersection-over-Union (IoU) loss in addition to the BCE loss for polyp segmentation, which improves model training.

\begin{table}
\vspace{-3mm}
\parbox{.45\linewidth}{
\caption{Quantitative comparison for stroke lesion segmentation.}\label{tab:stroke}
\vspace{-2mm}
\centering
\resizebox{0.9\linewidth}{!}{
\begin{tabular}{@{\hskip 1mm}ccccc@{\hskip 1mm}}
\toprule
Method  && DSC && IoU\\ 
\midrule
UNet~\cite{ronneberger2015u}  && 84.54 && 82.07 \\
TransUNet~\cite{chen2021transunet}  &&  87.37 && 83.14 \\
AttnUNet~\cite{oktay2018attention}  &&  85.30 && 83.28 \\
\midrule
SETR~\cite{zheng2021rethinking} && 82.88 && 77.40 \\
SegFormer~\cite{xie2021segformer}  && 81.45 && 79.56 \\
Swin Transformer~\cite{liu2021Swin}  && 84.74 && 80.73 \\
Patcher (Ours)  && \textbf{88.32} && \textbf{83.88} \\
\bottomrule
\end{tabular}
}
}
\hfill
\parbox{.45\linewidth}{
\caption{Quantitative comparison for polyp segmentation on Kvasir-SEG~\cite{jha2020kvasir}.}\label{tab:polyp}
\vspace{-2mm}
\centering
\resizebox{0.9\linewidth}{!}{
\begin{tabular}{@{\hskip 1mm}ccccc@{\hskip 1mm}}
\toprule
Method  && DSC && IoU\\ 
\midrule
UNet~\cite{ronneberger2015u}  && 78.89 && 67.81 \\
TransUNet~\cite{chen2021transunet}  &&  82.80 && 70.86\\
AttnUNet~\cite{oktay2018attention}  &&  77.08 && 61.48 \\
\midrule
SETR~\cite{zheng2021rethinking} && 75.30 && 61.60 \\
SegFormer~\cite{xie2021segformer}  && 87.39 && 78.81 \\
Swin Transformer~\cite{liu2021Swin}  && 85.58 && 77.64 \\
Patcher (Ours)  && \textbf{90.67} && \textbf{84.31}\\
\bottomrule
\end{tabular}
}
}
\vspace{-8mm}
\end{table}

\subsection{Comparison with State-of-the-Art Methods}
We compare our model, Patcher, with SOTA Transformer- and CNN-based segmentation models -- SETR~\cite{zheng2021rethinking}, SegFormer~\cite{xie2021segformer}, Swin Transformer~\cite{liu2021Swin}, U-Net~\cite{ronneberger2015u}, AttnUNet~\cite{oktay2018attention}, and TransUNet~\cite{chen2021transunet} -- using their released code. We use two common metrics for image segmentation -- dice score coefficient (DSC) and IoU.

\vspace{2mm}
\noindent\textbf{Quantitative Results.} 
The results for stroke lesion segmentation and polyp segmentation are shown in Tables~\ref{tab:stroke} and \ref{tab:polyp}, respectively. We can observe that Patcher outperforms the Transformer- and CNN-based baselines significantly. It is worth noting that the two segmentation tasks evaluate different aspects of the models. Stroke lesion segmentation requires the model to capture local details, and CNN-based models such as TransUnet~\cite{chen2021transunet} and AttnUNet~\cite{oktay2018attention} perform better than Transformer-based baselines. However, these CNN-based models perform much worse than Transformer-based models for polyp segmentation, \emph{e.g.,} 82.80 (TransUNet) \emph{vs.} 90.26 (Patcher), which is because polyp segmentation relies on better modeling of the global context where Transformers-based models excel. On both tasks, Patcher consistently outperforms the baselines, which shows its superior ability to model both local details and global context.

\begin{figure}[t!]
\includegraphics[width=\textwidth]{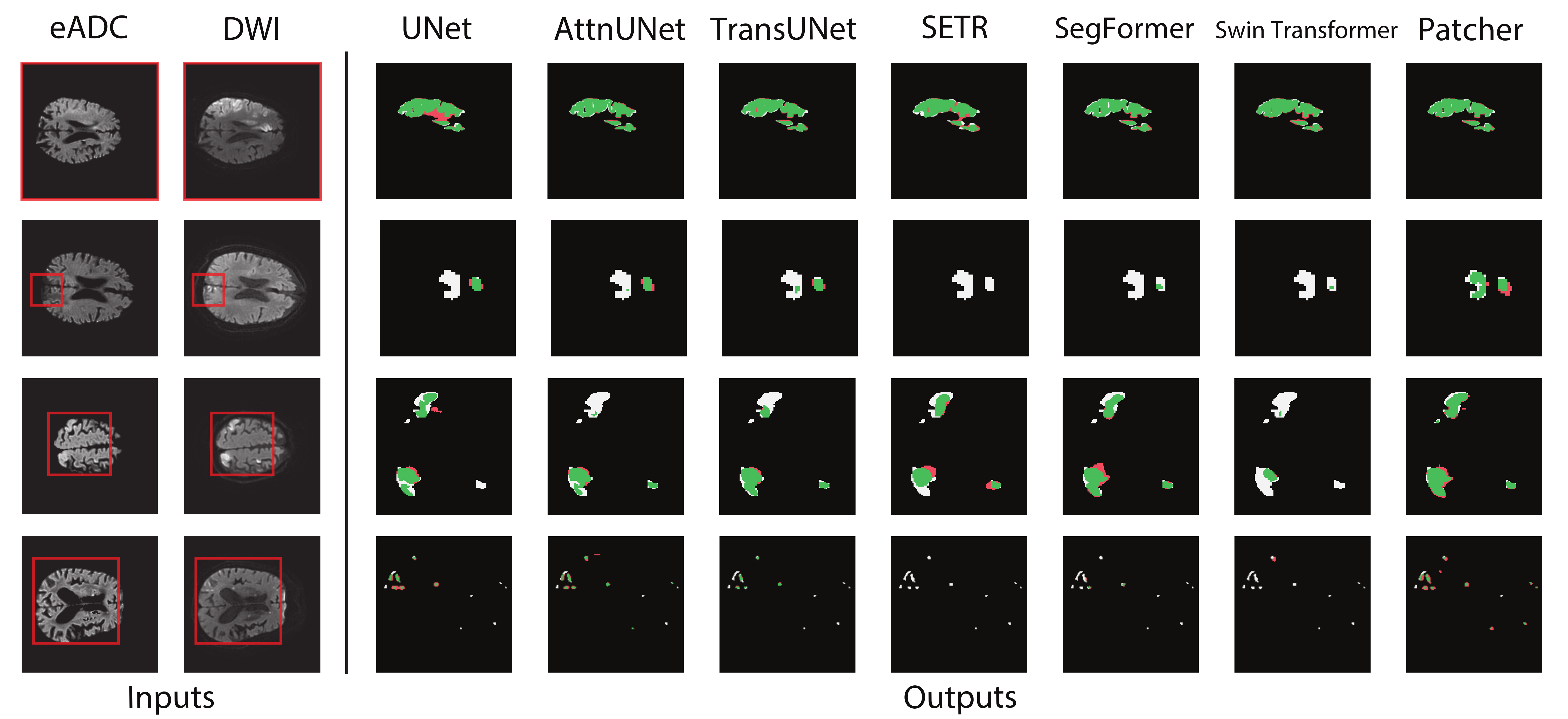}
\vspace{-7mm}
\caption{Visualization of stroke lesion segmentation. The outputs correspond to the red box. We highlight correct predictions (green), false positives (red), and false negatives (white). Patcher outputs more accurate segmentation especially for small lesions.}
\label{fig:results_stroke}
\vspace{-3mm}
\end{figure}

\begin{figure}[t!]
\includegraphics[width=\textwidth]{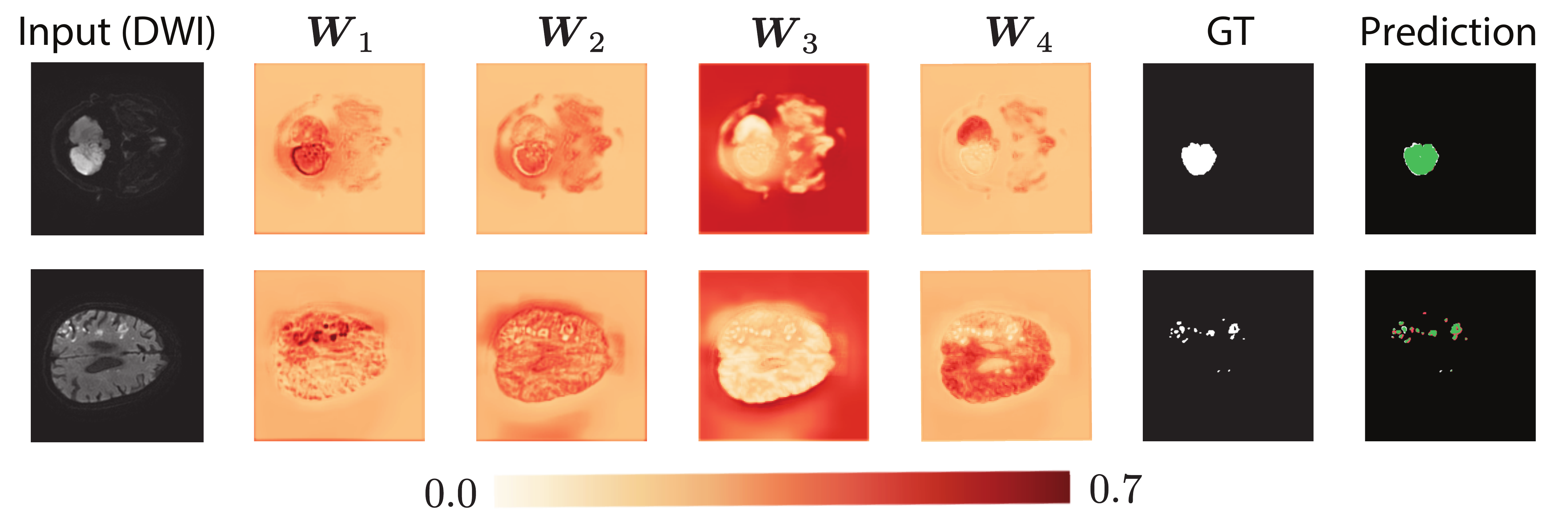}
\vspace{-7mm}
\caption{Visualization of four MoE weight maps $[\bs{W}_1, \bs{W}_2, \bs{W}_3, \bs{W}_4]$. Different weight maps focus on different areas (local details, global context, brain boundaries, \emph{etc.})}
\label{fig:moe_weights}
\vspace{-5mm}
\end{figure}

\vspace{2mm}
\noindent\textbf{Visual Results.} 
In Fig.~\ref{fig:results_stroke}, we provide a visual comparison of the segmentation maps from various models, where we paint correct predictions in green, false positives (FPs) in red, and false negatives (FNs) in white. We can observe that the baselines often have FPs and FNs, do not predict the segmentation boundaries well, and often miss small lesions. In contrast, Patcher has much fewer FPs and FNs, and can predict the segmentation boundaries accurately, even for very small lesions (Fig.~\ref{fig:results_stroke}, rows 2 and 4). This is due to Patcher's effective use of Transformers for capturing fine-grained pixel-level details. We also provide a visual comparison for polyp segmentation in Appendix~\ref{appx: polpy res}.

\vspace{2mm}
\noindent\textbf{Visualization of MoE weight maps.}
To better understand the MoE decoder, in Fig.~\ref{fig:moe_weights}, we visualize the MoE weight maps $[\bs{W}_1, \bs{W}_2, \bs{W}_3, \bs{W}_4]$ for the expert features for stroke lesion segmentation. We can see that MoE allows different weight maps to focus on different areas: $\bs{W}_1$ focuses on local details, $\bs{W}_2$ focuses on the global context inside the brain, $\bs{W}_3$ focuses on the boundaries of the brain, and $\bs{W}_4$ focuses on areas of the brain that are complementary to the lesions.

\begin{table}
\vspace{-3mm}
\parbox{.43\linewidth}{
\caption{Ablation studies of the Patcher encoder and MoE decoder.}
\label{tab:ablation_en_de}
\resizebox{0.95\linewidth}{!}
{
\begin{tabular}{@{\hskip 1mm}lccccccc@{\hskip 1mm}}
\toprule
Encoder && Decoder  && DSC && IoU\\ 
\midrule
SETR && MoE (Ours)  && 77.23 && 72.69 \\
Swin Transformer && MoE (Ours)  && 85.29 && 81.47 \\
\midrule
Patcher (Ours) && SETR-PUP   && 85.84 && 79.79 \\
Patcher (Ours) && U-Net  && 87.27 && 83.06 \\
\midrule
Patcher (Ours) && MoE (Ours)  && \textbf{88.32} && \textbf{83.88}\\
\bottomrule
\end{tabular}
}
}
\hfill
\parbox{.535\linewidth}{
\vspace{-0.2mm}
\caption{Effect of padded context size $P$ and large patch size $L$ for the 4 Patcher blocks.}\label{tab:ablation_size}
\vspace{-0.5mm}
\resizebox{\linewidth}{!}
{
\begin{tabular}{@{\hskip 1mm}cccccc@{\hskip 3mm}ccccccc@{\hskip 1mm}}
\toprule
\multicolumn{5}{c}{L = [32,32,32,32]} && \multicolumn{5}{c}{P = 8} \\
\cmidrule(l{0mm}r{0.5mm}){1-5} \cmidrule(l{0.3mm}r{0mm}){7-11}
Context $P$  && DSC && IoU && Large Patch $L$ && DSC && IoU\\ 
\cmidrule(l{0mm}r{0.5mm}){1-5} \cmidrule(l{0.3mm}r{0mm}){7-11}
0  && 86.85 && 83.40 && [64,64,64,32]  && 85.50 && 83.55\\
4 && 85.12 && 83.36 && [64,64,32,32]  && 87.70 && 83.78\\
8  && \textbf{88.32} && \textbf{83.88} && [32,32,32,32]  && \textbf{88.32} && \textbf{83.88}\\
16  && 86.88 && 83.31 && [32,16,16,16]  && 87.15 && 83.53\\
\bottomrule
\end{tabular}
}
}
\vspace{-10mm}
\end{table}

\subsection{Ablation Study}

We first perform ablation studies to evaluate the importance of the Patcher encoder and MoE decoder by replacing them with popular encoder or decoder designs. As shown in Table~\ref{tab:ablation_en_de}, when replacing the Patcher encoder with SETR~\cite{zheng2021rethinking} or Swing Transformer~\cite{liu2021Swin}, the performance decreases significantly. Similarly, when replacing the MoE decoder with SETR-PUP~\cite{zheng2021rethinking} and U-Net~\cite{ronneberger2015u}, the performs also drops considerably. This validates the importance of both designs.

We also conduct experiments to study the effect of varying the large patch size $L$ and padded context size $P$. As shown in Table~\ref{tab:ablation_size}, both $L$ and $P$ need to be carefully selected as they need to have enough context (not too small) while also focusing mainly on local features (not too large) to attain better performance.

\vspace{-2mm}
\section{Conclusions}
\vspace{-2mm}
In this paper, we proposed Patcher, a new Vision Transformer architecture for medical image segmentation that can extract fine-grained pixel-level features with Transformers only. By stacking a cascade of encoder blocks with increasing receptive fields, Patcher can extract both local and global features effectively. We also proposed a new MoE-based decoder that uses a gating network to select a suitable set of expert features from the encoder to output the prediction for each pixel. The use of MoE enables better specializations of the expert features and reduces interference between them. Extensive experiments and ablations validated the encoder and decoder design of Patcher and demonstrated that it outperforms SOTA Transformer- and CNN-based models substantially.

\bibliographystyle{splncs04}
\bibliography{ref}

\newpage
\appendix
\section{Details of the Transformer Blocks}
\label{appx: ViT}
\begin{figure}[h!]
\vspace{-7mm}
\centering
\includegraphics[width=0.8\textwidth]{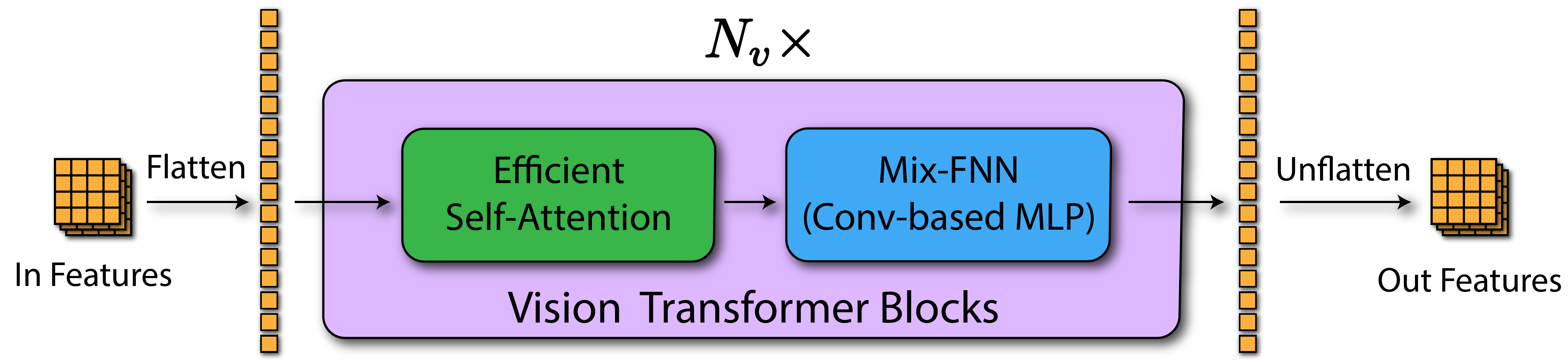}
\vspace{-4mm}
\caption{Details of the Transformer blocks used in the Patcher block (Fig. 2 of the main paper). Efficient self-attention and Mix-FNN are adopted from SegFormer.}
\label{fig:supp_transformer}
\vspace{-8mm}
\end{figure}

\section{Visualization for Polyp Segmentation }
\label{appx: polpy res}
\begin{figure}[h!]
\vspace{-7mm}
\centering
\includegraphics[width=0.95\textwidth]{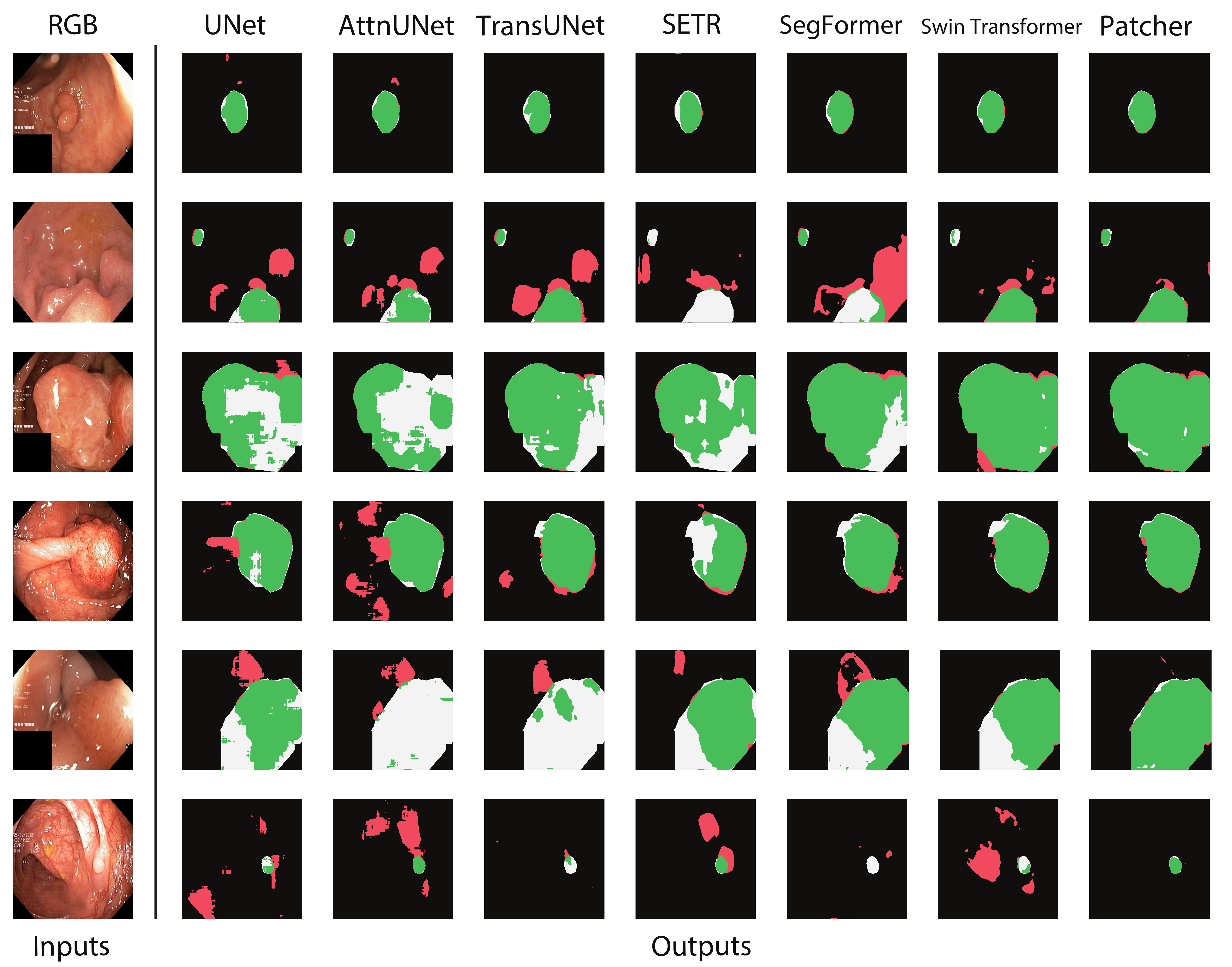}
\vspace{-3mm}
\caption{Visualization of polyp segmentation. We highlight correct predictions (green), false positives (red), and false negatives (white) in the red box of the input.}
\label{fig:supp_kvasir}
\vspace{-7mm}
\end{figure}

\newpage
\section{Additional Visualization for Stroke Lesion Segmentation }
\begin{figure}[h!]
\vspace{-6mm}
\centering
\includegraphics[width=\textwidth]{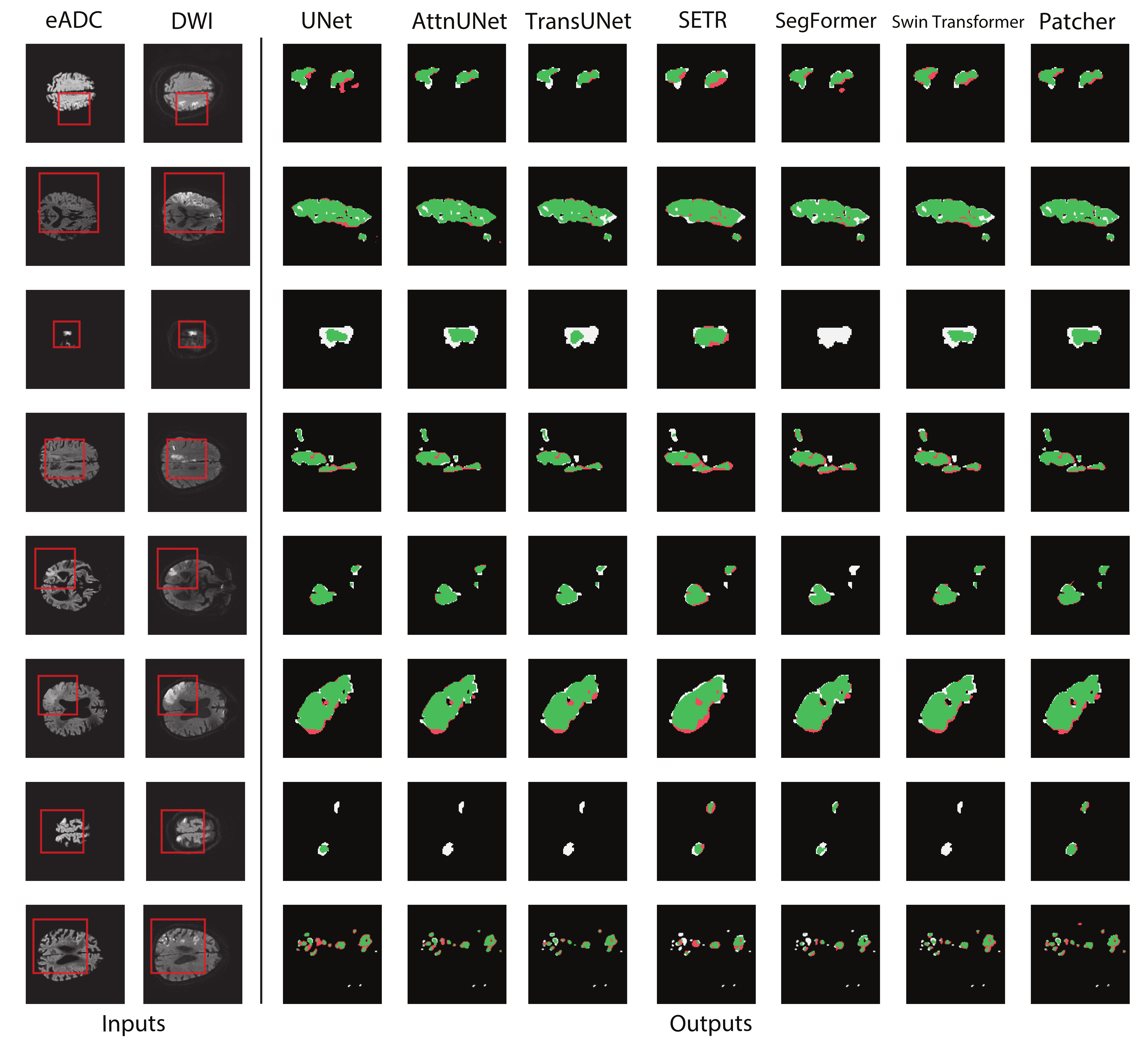}
\vspace{-6mm}
\caption{Additional visualization of stroke lesion segmentation. We highlight correct predictions (green), false positives (red), and false negatives (white) in the red box.}
\label{fig:supp_stroke}
\vspace{-8mm}
\end{figure}

\section{Additional Visualization of the MoE Weights}
\begin{figure}[h!]
\vspace{-10mm}
\centering
\includegraphics[width=\textwidth]{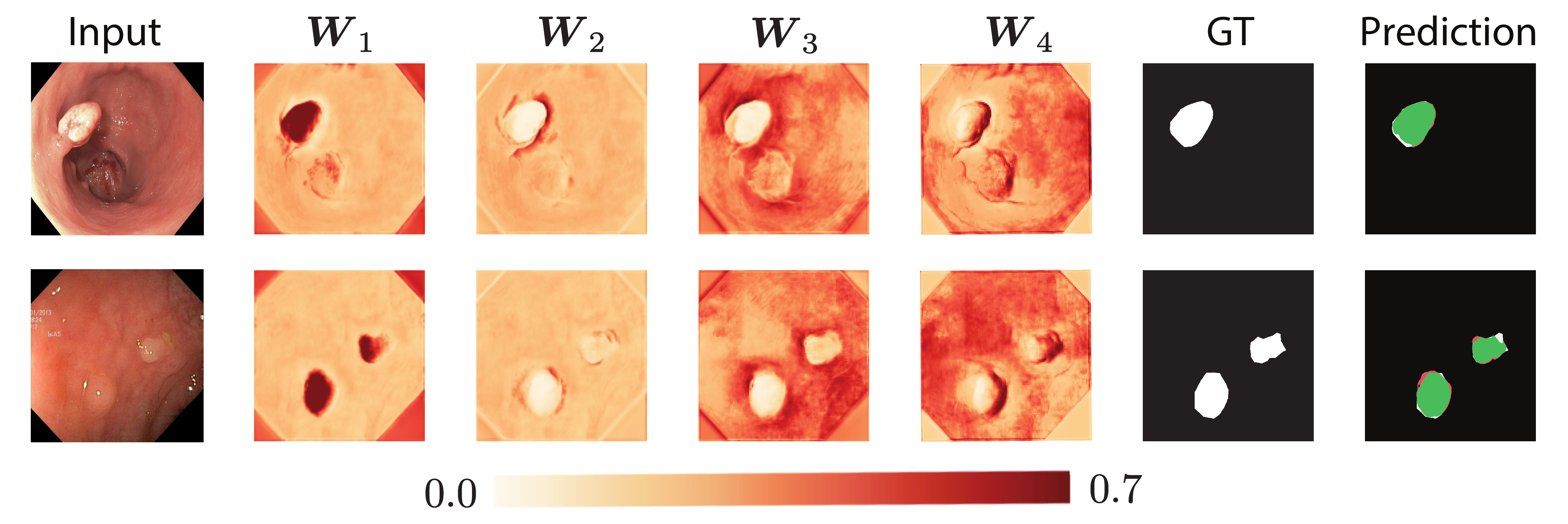}
\vspace{-6mm}
\caption{Additional visualization of four MoE weight maps $[\bs{W}_1, \bs{W}_2, \bs{W}_3, \bs{W}_4]$ for polyp segmentation. We can observe that different weight maps focus on different areas.}
\label{fig:supp_moe}
\vspace{-10mm}
\end{figure}

\end{document}